\documentclass{amsart}
\usepackage[mathscr]{eucal}
\usepackage{amscd}
\usepackage{amsfonts}
\usepackage{amsmath}
\usepackage{amsthm}
\usepackage{amssymb}
\usepackage{latexsym}

\begin{document}

\newcommand{\nc}{\newcommand}
\newcommand{\bom}{{_{\mathbf{\omega}}}}
\newcommand{\st}{\divideontimes}
\def\neweq{\setcounter{theorem}{0}}
\newtheorem{theorem}[]{Theorem}
\newtheorem{proposition}[]{Proposition}
\newtheorem{lemma}[]{Lemma}
\theoremstyle{definition}
\newtheorem{corollary}[]{Corollary}
\newtheorem{definition}[]{Definition}
\newtheorem{remark}[]{Remark}
\newtheorem{conjecture}[equation]{Conjecture}
\newtheorem{example}[theorem]{Example}
\newcommand{\dis}{{\displaystyle}}
\def\question{\noindent\textbf{Question.} }
\def\remark{\noindent\textbf{Remark.} }
\def\proof{\medskip\noindent {\textsl{Proof.} \ }}
\def\endproof{\hfill$\square$\medskip}
\def\str{\rule[-.2cm]{0cm}{0.7cm}}
\newcommand{\beq}{\begin{equation}\label}
\newcommand{\aand}{\quad{\text{\textsl{and}}\quad}}
\newcommand{\la}{\label}
\def\L{\mathcal{L}}
\def\D{\mathcal{D}}
\def\Ker{\mbox{\tt Ker}}
\def\Id{\mbox{\tt Id}}
\def\W{\mbox{\sl Wr}}
\def\R{\mathbb{R}}
\def\Z{\mathbb{Z}}
\def\vr{\varrho}
\def\vp{\varphi}
\def\tPsi{\tilde{\Psi}}
\def\tc{\tilde{c}}
\def\C{\mathbb{C}}
\def\c{\mathcal{C}}
\def\tU{\tilde{U}}
\def\tL{\tilde{L}}
\newcommand{\bk}{\boldsymbol{k}}
\title{Heat Kernel Coefficients for Two-dimensional Schr\"{o}dinger
Operators}
\author{Yuri Berest}
\address{Department of Mathematics,
Cornell University, Ithaca, NY 14853-4201, USA}
\email{berest@math.cornell.edu}
\thanks{Berest's work partially supported by NSF grant DMS 04-07502.}
\author{Tim Cramer}
\address{Department of Mathematics, Yale University, New Haven, CT 06520, USA}
\email{tim.cramer@yale.edu}
\author{Farkhod Eshmatov}
\address{Department of Mathematics, University of Michigan, Ann Arbor,
MI 48109-1043, USA}
\email{eshmatov@umich.edu}
\begin{abstract}
In this note, we compute the Hadamard coefficients of
(algebraically) integrable Schr\"odinger operators in two
dimensions. These operators first appeared in \cite{BL} and
\cite{B} in connection with Huygens' principle, and our result
completes, in a sense, the investigation initiated in those
papers.
\end{abstract}
\maketitle Let $\, \L = - \Delta_n + V\,$ be a Schr\"odinger
operator on $ \R^n $ ($ n \geq 1 $) with  $\c^\infty$-smooth
potential $\, V = V(x) \,$ defined in some open domain $\,\Omega
\subseteq \R^n $. Recall that the {\it heat kernel} of $\, \L \,$
is the solution $\, \Phi_{+}(x, \xi, t) \in \c^{\infty}(\Omega
\times \Omega \times \R^{1}_{+})\,$ of the initial value problem
\begin{equation}
\la{E1}
\bigg(\frac{\partial}{\partial t} + \L\bigg) \Phi_{+}(x,\xi,t) = 0\ ,\quad
\lim_{t \to 0+}\Phi_{+}(x,\xi, t)= \delta(x-\xi) \ ,
\end{equation}
where $\, \delta (x - \xi )\,$ is the Dirac delta-function on $
\Omega $ with support at $ \xi $. It is well known (see, e.g.,
\cite{R}, Sect. 3.2.1) that $ \Phi_{+}(x,\xi,t) $ has an
asymptotic expansion of the form
\begin{equation}
\la{E2}
 \Phi_{+}(x,\xi,t) \sim \frac{e^{-|x-\xi|^{2}/4t}}{(4\pi t)^{n/2}}
\bigg(1+ \sum_{\nu=1}^{\infty}U_{\nu}(x,\xi)\,t^{\nu} \bigg) \quad
\mbox{as}\ \ t \to 0{+} \ ,
\end{equation}
with coefficients $\, U_{\nu}(x, \xi) \in \c^\infty(\Omega \times
\Omega) \,$ determined by the following  {\it transport equations}
\begin{equation}
\la{E3}
(x-\xi,\partial_{x})\,U_{\nu}(x,\xi)+\nu\,U_{\nu}(x,\xi)= - \L
[U_{\nu-1}(\cdot,\xi)](x) \ , \quad \nu = 1,\,2, \, \ldots
\end{equation}
The system \eqref{E3} has a unique solution $\,\{U_{\nu}(x,
\xi)\}_{\nu = 0}^{\infty}\,$ if one sets $\, U_{0}(x, \xi) \equiv
1 \,$ and requires each $\, U_{\nu}(x,\xi)\,$ to be bounded in a
neighborhood of the diagonal $ x = \xi $. Following \cite{G}, we
will refer to $\,\{U_{\nu}(x, \xi)\}\,$ as the {\it Hadamard
coefficients} of the operator $\L $.

In general, calculating the Hadamard coefficients for a given
potential $\, V\,$ is a difficult problem. Of special interest are
potentials, for which the heat kernel expansion \eqref{E2} is
finite, i.~e. the sum in the right-hand side of \eqref{E2} has
only finitely many nonzero terms. In this case, formula \eqref{E2}
yields not only a ``short-time'' asymptotics, but an exact
analytic representation for $ \Phi_{+}(x,\xi,t) $ valid for all $
t \in \R^1_{+} $. Such potentials are usually called {\it Huygens'
potentials} in view of an important role they play in the theory
of Huygens' principle (see, e.~g., \cite{BV}, Ch.~I).

The problem of describing all Huygens potentials goes back to
Hadamard's classical treatise \cite{H} and still remains open in
all dimensions, except for $ n = 1 $ (see \cite{L}) and $ n = 2 $
(see \cite{B}). In dimension one, these potentials coincide with
the well-known Adler-Moser potentials \cite{AM}, which are the
rational solutions of the Korteweg-de Vries hierarchy of nonlinear
integrable PDE's (see \cite{S} or \cite{BV}, Ch.~3, Sect.~3.2);
apart from the original works \cite{LS} and \cite{L},
their Hadamard coefficients
have been studied recently in \cite{Gr}, \cite{I1}, \cite{I2}
and \cite{Ha}.

In the present paper, we will deal with the two-dimensional
Huygens potentials, which have received so far much less attention
(see, however, \cite{CFV}). Our main result (Theorem~\ref{T2})
provides simple and explicit formulas for the Hadamard
coefficients of the corresponding Schr\"odinger operators.
It is surprising that these formulas do not seem to have analogues
in dimension one: in a sense, the two-dimensional case is simpler
than the one-dimensional one!

We begin by recalling the main result of \cite{BL} and \cite{B}.
Let $\, k = (k_0,\,k_1,\,\ldots\,,k_m) \,$ be a finite, strictly
increasing sequence of integers, with $\,k_0 = 0\,$,
and let $ \varphi_i $ be real numbers given one for each integer
$\, k_i $, with $\,\vp_0=0\,$. Passing to the polar coordinates
$\, (x_1, \,x_2) = (r \cos \varphi,\, r\sin \varphi)\,$
in $ \R^2 $, we associate to these data the following potential
\begin{equation}
\la{E4} V_{k}(x_1, x_2) =  -
\frac{2}{r^2}\,\frac{\partial^2}{\partial \varphi^2}
\log \W \, [\chi_0,\, \chi_1,\, \ldots\, ,\, \chi_m]\ ,
\end{equation}
where $\,\chi_i:= \cos(k_i \varphi + \varphi_i)\,$, $\, i =
0,\,1,\,\ldots, m\,$, and $\, \W \, [\chi_0,\, \chi_1,\,\ldots\,
,\, \chi_m]  \,$ is the Wronskian of the set $\,\{\chi_i\}\,$
taken with respect to the variable $ \varphi $. As all $ k_i$'s
are integers, $\, V_k \,$ is a single-valued rational function on
$ \R^2 $, homogeneous of degree $ -2 $, whose analytic
continuation to $ \C^2 $ has singularities along certain lines
passing through the origin. For example, for $ k =
(0,\,1,\,3,\,4)\,$, with all $\, \vp_i$'s being $ 0 $, we have
$$
V_{k}(x_1, x_2) =
\frac{12\,(49\, x_1^4 + 28 \,x_1^2 \,x_2^2 - x_2^4)}{x_2^2\,(7 x_1^2 + x_2^2)^2}\ .
$$
In general, \eqref{E4} depends on both the choice of $ k_i$'s and
the choice of $\,\vp_i$'s, though we suppressed the latter from
our notation.
\begin{theorem}[\cite{BL, B}]
\la{T1} A (locally) smooth function $ V $ on $\R^2$, which is
homogeneous of degree $ -2 $, is a Huygens potential if and only
if $\, V = V_k(x_1, x_2)\,$ for some integer sequence $ k = (k_i)
$ and real numbers $ (\varphi_i) $.
\end{theorem}
\begin{remark}
\la{R1} The ``if'' part of Theorem~\ref{T1} was first proven in
\cite{BL} and then reproven by a different method, together with the
``only if'' part, in \cite{B} (see {\it loc. cit.}, Theorem~1.1).
The assumption that $ V $ is homogeneous can be relaxed and
replaced by a weaker condition that $ V $ and all the
Hadamard coefficients of $V$ are algebraic functions (see \cite{CFV}).
\end{remark}

\vspace{1ex}

The main result of this paper can now be encapsulated in
\begin{theorem}
\la{T2} The Hadamard coefficients of  $\, \L = -\Delta_2 + V_k \,$
with potential \eqref{E4} are given, in terms of polar coordinates
$\, x = (r \cos \vp,\, r\sin \vp)\,$ and $\, \xi = (\vr \cos
\phi,\, \vr \sin \phi)\,$, by
\begin{equation}
\la{E5} U_{\nu}(x,\xi) =  \frac{(-2)^\nu}{(r \vr)^\nu}
\sum_{i=0}^m\, c_i\,
\Psi_i(\vp)\,\Psi_i(\phi)\,T^{(\nu)}_{k_i}(\cos(\vp -\phi))\
,\quad \forall\,\nu \ge 0 \ ,
\end{equation}
where
\begin{equation}
\la{E6} \Psi_i := \frac{\W \, [\chi_0,\,\chi_1,\,
\ldots\,,\,\chi_{i-1}, \, \chi_{i+1},\,\ldots\,,\,\chi_m]}{\W \,
[\chi_0,\, \chi_1,\,\ldots\,, \chi_m]}\ ,\qquad c_i :=
\prod_{\substack{j=0\\j\not=i}}^m(k_i^2 - k_j^2)\ ,
\end{equation}
and $ T_N(z) := \cos(N \arccos z) $ is the $ N$-th Chebyshev
polynomial, with $\,T_N^{(\nu)}(z) \,$ being its derivative of
order $ \nu $ with respect to $\,z\,$.
\end{theorem}
\begin{remark}
It follows immediately from \eqref{E5} that $\,U_{\nu}(x, \xi) \equiv 0 \,$
for $\,\nu > k_m \,$, implying that $ V_k $ is a Huygens potential.
For $\, k \,$ of length one, i.~e. $\, k = (0,\, N)\,$, formula \eqref{E4} yields
the Calogero-Moser potential of dihedral type $ I_2(N) $, and in this special case
the coefficients \eqref{E5} have already been found in \cite{BL}
(see {\it loc. cit.}, Sect.~IV, Example~1).
\end{remark}

There are several ways to prove Theorem~\ref{T2}. Perhaps, the
most straightforward one is to use a differential recurrence
relation between the Hadamard coefficients of operators $\, \L_k
\,$ and $\, \L_{\tilde{k}} \,$, where $\, \tilde{k} = (k,\,
k_{m+1}) \,$ is obtained by adding one integer on top of $\,k\,$
(see \cite{BL}, (82)). This method requires double induction (in $
m $ and $ \nu $) and leads to rather unwieldy calculations.

Here, we will offer a more illuminating argument based on the
remarkable fact that the coefficients $\,U_{\nu}(x, \xi)\,$ appear
not only in  fundamental solutions of the heat equation but also
in its elliptic and hyperbolic counterparts\footnote{For an
excellent survey on this classical subject we refer the reader to
\cite{Ba}.}. Instead of the Cauchy problem \eqref{E1}, we will
consider
\begin{equation}
\la{E6'} \L [G(\,\cdot\,,\xi)](x) = \delta(x - \xi)\ , \quad
G(\,\cdot\,,\xi) \in \D'(\Omega)\ ,
\end{equation}
where $\,\D'(\Omega)\,$ denotes the space of distributions on
$\c^\infty$-functions with compact support in $\,\Omega\subseteq
\R^n \,$. Of course, unlike the heat kernel, $\,G(x,\xi)\,$ is not
uniquely determined by \eqref{E6'}, but only up to adding smooth
functions from $\, \Ker(\L) \,$. The problem is now to describe
the singularities of $\,G(x, \xi)\,$. In modern language, the
solution to this classical problem is given in terms of Riesz
distributions, and as in the hyperbolic case (see, e.g.,
\cite{D}), it depends on whether $\, n\,$ is even or odd.
Specifically, if $ n \geq 2 $ is even, $\,G(x, \xi)\,$ has the
following asymptotics ``in smoothness'' (cf. \cite{Ba}, Sect. 3,
(1.4)):
\begin{equation}
\la{E7} G(x,\xi) \sim \sum_{\nu = 0}^{\infty}
\tU_\nu(x,\xi)\,S_{\nu - \frac{n-2}{2}}'(\gamma)\ ,
\end{equation}
where $\, \gamma = |\,x - \xi\,|^2 \,$ is the square of the
Euclidean distance between $ x $ and $ \xi $, $\, S_{\lambda}(t)
:= t_{+}^{\lambda}/\Gamma(\lambda+1)\,$ is the family of
Riemann-Liouville distributions on $\R^1$, depending analytically
on the parameter $\,\lambda \in \C\,$, and $\, S_{\lambda}' := d
S_{\lambda}/d \lambda \,$ are the adjoint distributions of $
S_\lambda $ with respect to $ \lambda $ (see \cite{GS}, Ch.~VI,
Sect.~2). The  coefficients $\,\tU_\nu(x,\xi)\,$ in \eqref{E7} are
smooth functions, which satisfy, up to rescaling factor
$\,-1/4\,$, the {\it same} transport equations \eqref{E3} as the
heat kernel coefficients $\,U_\nu(x,\xi)\,$. Like in the heat
kernel case, this can be verified by direct calculation,
substituting \eqref{E7} into \eqref{E6'}. By uniqueness of the
regular solution of \eqref{E3}, we thus have
\begin{equation}
\la{E8}
 \tU_\nu(x, \xi) = \left(-\frac{1}{4}\right)^{\!\nu}
U_\nu(x, \xi)\quad \mbox{for all}\ \nu \geq 0\ .
\end{equation}

Now, let $\,n=2\,$ and assume that $ V = V(x_1,\,x_2) $
is (locally) analytic, as are our potentials \eqref{E4}.
In this case, the distributions $\,S_{\lambda}' \,$
appear in \eqref{E7} only with non-negative integer $ \lambda$'s,
and for such $ \lambda$'s they can be easily calculated
\begin{equation}
\la{E9} S_{\nu}'(t) := \left.\frac{d S_\lambda(t)}{d \lambda}\,
\right|_{\lambda = \nu} = \frac{1}{\nu!}\,t^\nu_{+}\log t + C_\nu\,
t^\nu_{+} \ ,\quad \nu = 0,\,1,\,2,\,\ldots\ ,
\end{equation}
where $\,C_\nu \in \R \,$ are some constants, with $ C_0 = 0 $.
Substituting \eqref{E8} and \eqref{E9} in \eqref{E7}, we get
$$
G(x, \xi) \sim W(x,\xi)\,\log \gamma + \, \ldots\ ,
$$
where ``$\,\ldots \,$'' denote smooth functions, which do not
contribute to singularities of $ G(x, \xi)$, and
\begin{equation}
\la{E10} W(x,\xi) := \sum_{\nu = 0}^\infty \,
\frac{1}{(-4)^\nu \nu!}\,U_\nu(x,\xi) \,\gamma^\nu\ .
\end{equation}
Since $ V $ is analytic, the series \eqref{E10} uniformly
converges in a neighborhood of the diagonal $\, x = \xi \,$. This
was already observed by Hadamard, who called $ W(x,\xi) $ the {\it
logarithmic term} of the ``elementary solution''  $ G(x,\xi) $.
He also showed that $ W(x, \xi) $ is the (unique) analytic
solution of the Goursat problem
\begin{equation}
\la{E11}
\L[W(\,\cdot\,,\xi)](x) = 0 \ ,\quad W|_{\gamma = 0}\, = \, 1\ ,
\end{equation}
with the boundary condition given on the (complex) characteristics
of $ \L $ (see \cite{H}, Ch.~III, Sect.~46).

Now, to prove Theorem~\ref{T2} we will simply solve \eqref{E11}
for the operator $ \L $ with potential \eqref{E4}, using
the method of separation of variables, and then recover the
coefficients $ U_\nu(x, \xi) $ by expanding the solution in the
 vicinity of
$\, \gamma = 0 \,$ as in \eqref{E10}.

First, we introduce a one-dimensional Schr\"odinger operator $ L $
by writing $ \L $ in terms of polar coordinates (cf. \cite{B},
Sect.~3):
\begin{equation}
\la{E22} \L = - \frac{\partial^2}{\partial r^2} - \frac{1}{r}
\frac{\partial}{\partial r} + \frac{1}{r^2}\, L \ .
\end{equation}
Explicitly, $\, L = -\partial^2\!/\partial \vp^2 + V_k(\cos \vp,
\,\sin \vp) \,$, where $ V_k $ is given by \eqref{E4}. Next, we
prove
\begin{lemma}
\la{L1} If $\, \Psi = \Psi(\varphi) \,$  satisfies $\, L[\Psi] = k^2 \Psi \,$
with some integer $\, k \ge 0 $, then
$$
\Psi(\varphi)\, T_k [\,(r/\vr + \vr/r)/2\,] \,\in\, \Ker(\L)\quad
\mbox{for any}\quad  \vr \not= 0\ .
$$
\end{lemma}
\begin{proof}
Changing the variables $\,  (r, \,\varphi) \mapsto (z,
\,\varphi)\,$ in \eqref{E22}, with $\, z := (r/\vr + \vr/r)/2 \,$,
yields
$$
\L =  \frac{1}{r^2} \left[(1-z^2)\frac{\partial^2}{\partial z^2}
-z \frac{\partial}{\partial z} + L \right]\ .
$$
Since $\, T_k(z) \,$ is an eigenfunction of $\,(1-z^2)d^2\! /dz^2
- z d/dz\,$ with eigenvalue $ - k^2 $, the claim is obvious.
\end{proof}
\begin{lemma}
\la{L2} The functions $\, \Psi_i \,$ defined by \eqref{E6} satisfy
\begin{equation}
\la{E12} L[\Psi_i] = k^2_i \,\Psi_i\ , \quad i =
0,\,1,\,\ldots\,,\, m \ ,
\end{equation}
\begin{equation}
\la{E13} \sum_{i=0}^m c_i\,\Psi_i(\varphi)\,\Psi_{i}(\phi)\,\cos(k_i
(\varphi - \phi)) = 1\ .
\end{equation}
\end{lemma}
\begin{proof} The equations \eqref{E12} simply say that $ \Psi_i$'s
are
eigenfunctions of $ L $ with eigenvalues $ k_i^2 $. This follows
directly from Crum's classical theorem (see \cite{C}, p.~124).

The identity \eqref{E13} seems more interesting: we could not find
it in the literature, so we will prove it in detail by
induction on $m$.

The case $\, m = 1 \,$ is straightforward. Writing $ S_m(\vp,
\phi) $ for the left-hand side of \eqref{E13}, we now fix $ m \ge
1 $ and assume that $ S_m(\vp, \phi) \equiv 1 $ for all sequences
of integers $\, (k_0 = 0,\, k_1,\, \ldots,\, k_m) \,$ and reals $\,(\vp_0 = 0, \,
\vp_1, \, \ldots ,\,\vp_m)\,$ of length $\,m\,$. To make the induction
step we add $\, k_{m+1} \in \Z \,$, $\, k_{m+1} > k_m \,$, and $
\vp_{m+1} \in \R $ to these sequences and consider
\begin{equation}
\la{Sm}
S_{m+1}(\vp,\phi) := \sum_{i=0}^{m+1} \tc_i \,\tPsi_i(\vp)\, \tPsi_i(\phi)
\cos(k_i (\vp - \phi))\ ,
\end{equation}
where $ \tPsi_i $ and $ \tc_i $ are defined by formulas \eqref{E6}
with $ m $ replaced by $ m+1 $. By Crum's Theorem, $\, \tPsi_i $'s
are eigenfunctions of an operator $ \tL $ obtained from $ L $ by
applying the Darboux transformation (cf. \cite{B}, (3.29)):
\begin{equation}
\la{Dar}
L = A_m^* \circ A_m + k_{m+1}^2 \ \mapsto \
\tL = A_m \circ A_m^* + k_{m+1}^2\ ,
\end{equation}
where
\begin{equation}
\la{sh}
A_m := \tPsi_{m+1}^{-1}\circ \left(\frac{\partial}{\partial \vp}\right)
\circ\tPsi_{m+1}\ , \quad
A_m^* := - \tPsi_{m+1}\circ \left(\frac{\partial}{\partial \vp}\right)\circ
\tPsi_{m+1}^{-1}\ .
\end{equation}
It is immediate from \eqref{sh} that $\, A_m^*[\tPsi_{m+1}] =
0\,$. On the other hand, we have
\begin{equation}
\la{Id1}
A_m^*[\tPsi_i] = - \Psi_i\quad \mbox{for all}\ \, i = 0,\,\ldots, m\ .
\end{equation}
In fact, a trivial calculation shows that \eqref{Id1} is
equivalent to
\begin{equation}
\la{Idd1}
\frac{\partial}{\partial \vp}\left(\frac{W_i}{W_{m+1}}\right) =
\frac{W\,W_{i, m+1}}{W_{m+1}^2}\ ,
\end{equation}
where $ W $ denotes the Wronskian of the set $ \{\chi_0,
\chi_1,\ldots, \chi_{m+1}\} $, and $ W_i\,$,$\, W_{i,m+1} $ are
the Wronskians of this set with functions $\, \chi_i\,$ and $\,
\{\chi_i,\,\chi_{m+1}\}\,$ being omitted. Now, to prove
\eqref{Idd1} consider the system of linear equations
\begin{equation}
\la{idd3}
\sum_{i=0}^{m} \chi_{i}^{(k)} y_i = \chi_{m+1}^{(k)}\ ,\quad
k = 0,\,1,\,\ldots,\,m\ .
\end{equation}
By Cramer's Rule, the solution to this system is given by $\,y_i =
(-1)^{m+i}\,W_i/W_{m+1} \,$. On the other hand, differentiating
both sides of \eqref{idd3} with respect to $ \vp $ yields $\,
\sum_{i=0}^m \chi_i^{(k)}\,y_i' = 0\,$ for $\, k
=0,\,1,\,\ldots,\,m-1\,$ and $\, \sum_{i=0}^m \chi_i^{(m)}\,y_i' =
W/W_{m+1}\,$ for $\, k = m \,$. Solving these equations for $ y_i'
\,$, we get $\, y_i' = (-1)^{m+i} W\,W_{i,m+1}/W_{m+1}^2 \,$,
which is equivalent to \eqref{Idd1}.

It follows from \eqref{Id1} that $\, A_m[\Psi_i] = - A_m
A_m^*[\tPsi_i] = - (\tL - k_{m+1}^2 )\tPsi_i\,$. Hence
\begin{equation}
\la{Id2}
A_m[\Psi_i]  =
(k_{m+1}^2 - k_i^2) \tPsi_i \quad \mbox{for all}\ \, i = 0,\,\ldots, m\ .
\end{equation}
Now, applying $ A_{m} $ and $ A_m^* $ to \eqref{Sm}, we can
formally relate $ S_{m+1}(\vp, \phi) $ to $ S_{m}(\vp, \phi) $. In
fact, a straightforward calculation using \eqref{Id1} and
\eqref{Id2} shows
\begin{equation}
\la{diff}
A^{*}_{m}[S_{m+1}(\,\cdot\,,\,\phi)] + A^{*}_{m}[S_{m+1}(\vp,\,\cdot\,)]
= A_m[S_m(\,\cdot\,,\,\phi)] + A_m[S_m(\vp,\,\cdot\,)] \ ,
\end{equation}
where omitted are the variables on which the differential
operators act. Since $\, S_m(\vp, \phi) \equiv 1 \,$ by induction
assumption, we can regard \eqref{diff} as a first order PDE for
the function $ S_{m+1}(\vp, \phi) $. Specifically, substituting $
S_m = 1 $ into \eqref{diff} yields
\begin{equation}
\la{pde}
- \left(\frac{\partial}{\partial \vp} + \frac{\partial}{\partial \phi}
 \right) S_{m+1} + \left(\frac{\partial B}{\partial \vp} +
 \frac{\partial B}{\partial \phi}
 \right)\, S_{m+1} = \frac{\partial B}{\partial \vp} +
 \frac{\partial B}{\partial \phi}\ ,
\end{equation}
where $\, B := \log \,[\tPsi_{m+1}(\vp)\tPsi_{m+1}(\phi)] \,$. This
PDE can be easily integrated: changing the variables $\, x := (\vp
+ \phi)/2\,$, $\, t := (\vp-\phi)/2\,$ and $\, F :=
(S_{m+1}-1)\,e^{-B} \,$ transforms \eqref{pde} to the equation
$\,\partial F/\partial x = 0\,$, which means that $\,F\,$ depends
only on $ t $. It follows that
\begin{equation}
\la{F}
S_{m+1}(\vp, \phi) = 1 + F(\vp -\phi) \,\tPsi_{m+1}(\vp)\,\tPsi_{m+1}(\phi)\ ,
\end{equation}
where $ F $ is a differentiable function of one variable defined
on $ (0, \, 2\pi) $.

Now, if we replace $ k_m $ with $ k_{m+1} $ in the sum $\, S_m \,$
and repeat the above argument, adding $ k_m $ to the partition $\,
(k_0 = 0,\, k_1,\, \ldots,\, k_{m-1}, k_{m+1}) \,$, then, instead
of \eqref{F}, we get
\begin{equation}
\la{G} S_{m+1}(\vp, \phi) = 1 + G(\vp -\phi)
\,\tPsi_{m}(\vp)\,\tPsi_{m}(\phi)\ ,
\end{equation}
where $ G $ is another differentiable function on $ (0,\,2 \pi) $.
Comparing \eqref{F} and \eqref{G} shows that $\, F = G \equiv 0
$. Indeed, if one of these functions ($ F $ say) is nonzero, then
by continuity $ F(\vp -\phi) \not= 0 $ in an open subset of $\,
(0, \, 2\pi) \times (0, \, 2\pi) \,$, and in that subset we have
\begin{equation}
\la{GF}
{G(\vp - \phi)}/{F(\vp -\phi)} = h(\vp)\,h(\phi)\ ,
\end{equation}
where $\,h := \tPsi_{m+1}/\tPsi_{m}\,$. Differentiating both sides
of \eqref{GF} with respect to $ \vp $ and $\phi$ and adding the
results yields $\, h'(\vp)\,h(\phi) + h(\vp)\,h'(\phi) = 0\,$.
Whence, letting $\, \vp = \phi \,$, we see that $\, h \,$ must be
constant. This means that $ \tPsi_{m+1} $ is a multiple of $
\tPsi_m $, which is impossible, since $ \tPsi_{m} $ and $
\tPsi_{m+1} $ are nonzero eigenfunctions of $ \tL $ corresponding
to  different eigenvalues ($k_{m}^2$ and $ k_{m+1}^2 $
respectively). Thus $\, F \equiv 0 \,$, and therefore $\,
S_{m+1}(\vp, \phi) \equiv 1 $, finishing the induction.
\end{proof}

Now, combining the results of Lemmas~\ref{L1} and \ref{L2}, we see
at once that
\begin{equation}
\la{E14} W := \sum_{i=0}^m
\,c_i\,\Psi_i(\varphi)\,\Psi_i(\phi)\, T_{k_{i}}[\,(r/\vr + \vr/r)/2 \,]
\end{equation}
is a solution to \eqref{E11}. Indeed, by \eqref{E12} and
Lemma~\ref{L1}, each summand of \eqref{E14} lies in the kernel of
$ \L $, and hence $\,\L[W] = 0 \,$ by linearity of $ \L $. On the
other hand, in polar coordinates $\, \gamma = r^2 + \vr^2 - 2\, r
\vr \cos(\vp -\phi) \,$, so
\begin{equation}
\la{E55}
\frac{1}{2}\left(\frac{r}{\vr} + \frac{\vr}{r}\right) =
\frac{\gamma}{2\,r \varrho} +
\cos(\varphi-\phi)\ ,
\end{equation}
and therefore
$$
T_{k_{i}}[\,(r/\vr + \vr/r)/2 \,]\, =\, \cos(k_i(\varphi - \phi))
\quad \mbox{on} \quad \gamma = 0
$$
for all $ \,i = 0,\,1,\,\ldots\,, m \,$. The boundary condition for
$\,W\,$ follows then from \eqref{E13}.

Now, by \cite{B}, Lemma~3.1, $\, U_\nu(x,\xi) \,$ are homogeneous
functions of $x$ and $\xi$, which can be written in terms of the polar
coordinates as
\begin{equation}
\la{E66}
U_\nu(x,\xi) = \frac{1}{(r \vr)^\nu}\,\sigma_{\nu}(\vp,\,\phi)\ ,
\quad\,\nu \ge 0\ .
\end{equation}
Clearly, there is at most one expansion of $ W(x, \xi) $ of the
form \eqref{E10} with coefficients \eqref{E66}. To find this
expansion, we take the obvious Taylor formulas (see
\eqref{E55})
\begin{equation}
\la{E88}
T_{k_i}[\,(r/\vr + \vr/r)/2 \,] = \sum_{\nu=0}^{\infty}
\,\frac{1}{\nu !} \,\left(\frac{\gamma}{2\,r \varrho}\right)^\nu\,
T^{(\nu)}_{k_{i}}(\cos(\varphi - \phi))\ ,
\end{equation}
substitute \eqref{E88} into \eqref{E14} and reorder summations.
Collecting coefficients under the different powers of $ \gamma $
gives then the desired formulas \eqref{E5}.

In the end, we note that our operators $ \L $ are examples of
algebraically integrable Schr\"odinger operators (see \cite{CV}).
Such Schr\"odinger operators possess special eigenfunctions
called the {\it Baker-Akhiezer functions}. Like fundamental
solutions above, the Baker-Akhiezer functions have asymptotic
expansions, with coefficients satisfying (up to rescaling factor $1/2$)
the same transport equations \eqref{E3}. This was originally
discovered in the special case of
Calogero-Moser potentials in \cite{BV1}, but the argument of
\cite{BV1} applies to any homogeneous operator (see, e.g.,
\cite{CFV}). In combination with Theorem~\ref{T2}, this yields
\begin{theorem}
\la{T3}
The Baker-Akhiezer function of
$\, \L = - \Delta_2 + V_k(x_1, x_2) \,$  is given by
$$
\Psi_{\sf BA}(x,\,\xi) = \left(\,\sum_{\nu = 0}^{k_m}
\,\frac{1}{2^\nu}\,U_\nu(x,\,\xi)\,\right) \,e^{(x,\,\xi)}\ ,
$$
where $\, U_{\nu}(x,\,\xi) \,$ are the same coefficients as in
\eqref{E5}.
\end{theorem}
\bibliographystyle{amsalpha}

\end{document}